\documentclass[a4paper,pra,floatfix,twocolumn]{revtex4-1}

\usepackage[T1]{fontenc}
\usepackage[latin9]{inputenc}
\usepackage{color}
\usepackage{amsmath}
\usepackage{amssymb}
\usepackage{graphicx}
\usepackage{esint}

\makeatletter

\providecommand{\tabularnewline}{\\}

\@ifundefined{textcolor}{}
{%
 \definecolor{BLACK}{gray}{0}
 \definecolor{WHITE}{gray}{1}
 \definecolor{RED}{rgb}{1,0,0}
 \definecolor{GREEN}{rgb}{0,1,0}
 \definecolor{BLUE}{rgb}{0,0,1}
 \definecolor{CYAN}{cmyk}{1,0,0,0}
 \definecolor{MAGENTA}{cmyk}{0,1,0,0}
 \definecolor{YELLOW}{cmyk}{0,0,1,0}
}

\@ifundefined{definecolor}
 {\usepackage{color}}{}

\global\long\def\ket#1{\left| #1\right\rangle }

\global\long\def\bra#1{\left\langle #1 \right|}

\global\long\def\av#1{\left\langle #1 \right\rangle }

\global\long\def\tr{\text{Tr}}

\global\long\def\abs#1{\left|#1\right|}

\newcommand{\be}{\begin{eqnarray}}\newcommand{\ee}{\end{eqnarray}}\def\beq{\begin{equation}}\def\eeq{\end{equation}}

\makeatother

\begin{document}

\title{Strongly interacting bosons in multi-chromatic potentials supporting mobility edges:
 localization, quasi-condensation and expansion dynamics}



\author{Pedro Ribeiro}

\author{Masudul Haque}

\author{Achilleas Lazarides}

\affiliation{Max-Planck-Institut f\"{u}r Physik 
komplexer Systeme, N\"{o}thnitzer Stra{\ss}e 38, D-01187 Dresden, Germany}

\begin{abstract}

We provide an account of the static and dynamic properties of hard-core bosons in a one-dimensional
lattice subject to a multi-chromatic quasiperiodic potential for which the single-particle spectrum
has mobility edges.  We use the mapping from strongly interacting bosons to weakly interacting
fermions, and provide exact numerical results for hard-core bosons in and out of equilibrium.  In
equilibrium, we find that the system behaves like a quasi-condensate (insulator) depending on
whether the Fermi surface of the corresponding fermionic system lies in a spectral region where the
single-particle states are delocalized (localized).  We also study non-equilibrium expansion
dynamics of initially trapped bosons, and demonstrate that the extent of partial localization is
determined by the single-particle spectrum.

\end{abstract}

\maketitle

\section{Introduction \label{sec_intro}}

The localization of quantum particles and waves in disordered media has been a key theme common to
several fields of physics since Anderson's prediction of the effect half a century ago
\cite{Anderson_1958}.  Among the different setups leading to localization effects, quasiperiodic
potentials with incommensurate periods \cite{Aubry_1980} provide a particularly attractive
realization, as their mathematical simplicity allows for various exact results.  A one-dimensional
(1D) system subject to such a quasiperiodic potential (Aubry-Andre potential \cite{Aubry_1980}) has
a localization transition at a finite value of the potential strength, in contrast to a random
potential which in 1D causes localization at infinitesimal strength.  Recently, pioneering
experiments have explored the physics of bosons \cite{Roati_2008, Inguscio2, Inguscio3, Inguscio4}
and light \cite{Lahini_2009, Kraus_2012} in such potentials.

The single-particle spectrum acquires additional structure upon modification of the quasiperiodic
superlattice potentials e.g., when there are two superlattice potentials with different wavelengths
each incommensurate with the lattice, or by other modifications of the basic Aubry-Andre (AA)
potential \cite{Soukoulis_1982, Riklund_1986, Dassarma_1986, Sarma_1988, Dassarma_1988b,
  Fishman_1988, Hiramoto, Varga_1992, Scarola_2006, Boers_2007, Biddle_2009, Biddle_2010,
  Biddle_2011}.  In these cases, different energy regions of the single-particle spectrum might
become localized at different strengths of the potential, so that at some strengths of the potential
there are both localized and delocalized eigenstates.  Such a structure is known as a \emph{mobility
  edge} \cite{Soukoulis_1982}.  Mobility edges are well-known to exist in truly random potentials,
but only in higher dimension.

The interplay between disorder and interactions is a prominent theme in the study of strongly
correlated systems.
With the realization of quasiperiodic potentials hosting bosonic atoms \cite{Roati_2008, Inguscio2,
Inguscio3, Inguscio4}, a natural question is the behavior of interacting bosons in quasiperiodic
potentials.
Refs.\ \cite{Roux_2008,Deng_2008,Deng_2009,Orso_2009,Schmitt_2009,Zakrzewski_2009,Sanchez-Palencia_2010,Roscilde_2010,Shrestha_2010,Kai_2012,Cai1,Cai2,Cai3,Cai4,Cetoli_2010,Nessi_2011,Flach_2012}
have studied the Bose-Hubbard model in the AA potential for nonzero interactions.  Some of these
works \cite{Cai1,Cai2,Cai3,Cai4,Nessi_2011} have used the infinite-interaction or hard-core limit of
the Bose-Hubbard model, where multiple occupancies are disallowed.  The hard-core boson (HCB) model
can be mapped onto free fermions, allowing numerically exact calculations for relatively large
system sizes even in the absence of translation symmetry.
Ground state, finite-temperature, and non-equilibrium properties of hard-core bosons in the AA
potential have been analyzed in some detail in these studies \cite{Cai1,Cai2,Cai3,Cai4,Nessi_2011}.

In this article, we address ground-state and non-equilibrium properties of hard-core bosons in
multi-chromatic quasiperiodic potentials (extended Aubry-Andre models) where the single-particle
spectrum displays mobility edges.  Our main result concerning equilibrium properties is a connection
between the behavior of the many-body system (insulating or quasi-condensate) and the location of
the Fermi energy of the corresponding free-fermion system (which is the chemical potential for both
the corresponding free-fermion system and for the HCB).
The system acts like a quasi-condensate (or insulator) when the chemical potential is in an energy
region where the single-particle states are extended (or localized), irrespective of the nature of
single-particle states lying below the chemical potential.
This phenomenon is particularly puzzling in the case where the chemical potential is in a localized
region but there are filled extended states at lower energies.  One would expect superfluid
properties due to extended states being occupied.
We demonstrate this remarkable property, that the system behavior is determined by the location of
the chemical potential, through the study of various quantities (quasi-condensate fraction scaling,
off-diagonal order, entanglement entropy, etc).

We also study the dynamics after release of an initially trapped cloud of hard-core bosons in
multi-chromatic potentials.  Expansion of initially confined many-body systems in potentials
displaying mobility edges (higher-dimensional random potential) has been explored recently in two
experiments \cite{Kondov_2011, Jendrzejewski_2012}, providing motivation for theoretical
non-equilibrium calculations such as ours.

We will start in Section \ref{sec_model_method} with a description of the model and some
single-particle properties of multi-chromatic potentials.  Section \ref{sec_equilibrium} presents
various ground state properties, highlighting the connection to the position of the chemical
potential.  Section \ref{sec_dynamics} discusses the non-equilibrium expansion dynamics.

\section{Model and Method  \label{sec_model_method}}

\begin{figure}[tb]
\centering \includegraphics[width=0.8\columnwidth]{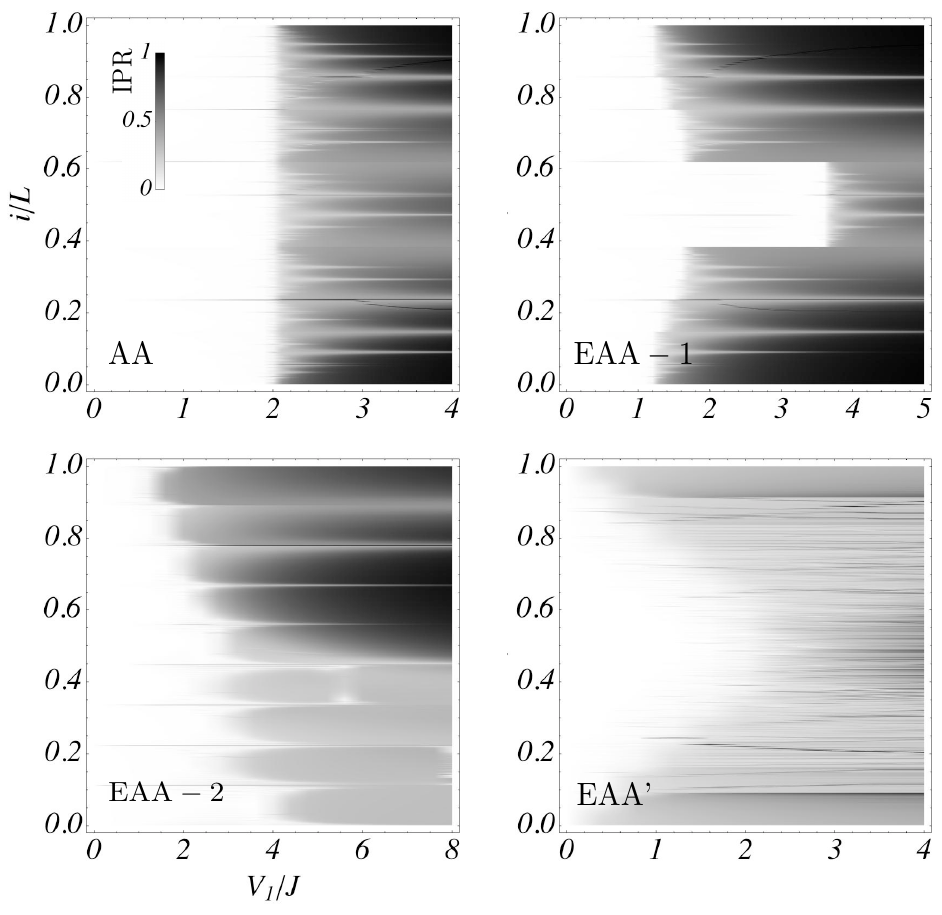}
\caption{  \label{fig:1-1} 
Inverse participation ratio of all single-particle eigenstates, for
the Aubry-Andre (AA) model and for the extended models EAA-1, EAA-2
and EAA'.  Here $i$ is the eigenstate index in order of increasing
eigenenergies.  Except for AA, at intermediate potential strengths
$V_{1}$ both localized and delocalized eigenstates coexist.  Mobility
edges are observed as a function of energy as one crosses  boundaries
between localized and extended regions.
}
\end{figure}

\begin{figure}[tb]
\centering \includegraphics[width=1\columnwidth]{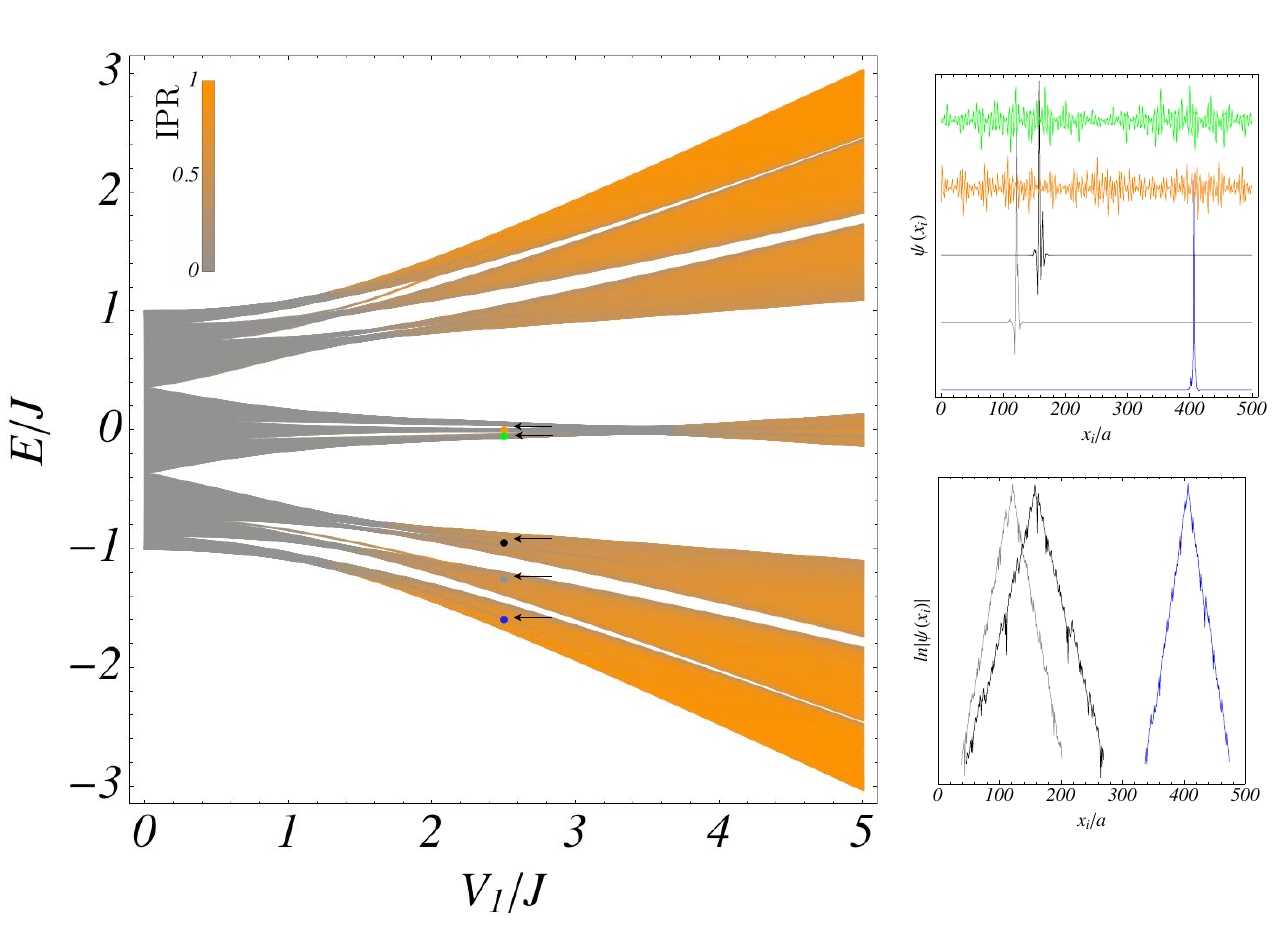}
\caption{\label{fig:1-2}
Left: Single-particle spectrum as a function of the quasi-periodic potential amplitude for the EAA-1
model.  The color code indicates IPR values.  The localization-delocalization transition occurs
where many eigenstates bunch together; this bunching occurs at different $V_1$ for different parts
of the spectrum.
Upper right: some representative wave functions for $V_1=2.5J$ (mobility edge region) for the states
indicated with colored dots in left panel.  Both localized and extended eigenfunctions are seen.
Lower right: absolute value of the localized wave functions is plotted in log-linear scale showing
that localization is exponential.
}
\end{figure}

We consider a system of $N_b$ hard-core bosonic atoms on a chain with $L$
sites and open boundary conditions, described by the  Hamiltonian
\begin{equation}
H  ~=~  -J\sum_{n}\left(b_{n}^{\dagger}b_{n+1}+b_{n+1}^{\dagger}b_{n}\right)
 ~+~ \sum_{n} V(n)b_{n}^{\dagger}b_{n}
\label{eq:hamiltonian-definition}
\end{equation}
where $b_{n}$ and $b_{n}^{\dagger}$ are bosonic creation and
annihilation operators,
$\left[b_{n},b_{n'}^{\dagger}\right]=\delta_{n,n'}$.  The hard-core
constraint is imposed through $b_{n}^{2}=b_{n}^{\dagger2}=0$. 
We use several quasi-periodic potentials, $V(n)$, specified below.

Such a hard-core bosonic system may be regarded as the limit of the
Bose-Hubbard model where the on-site Bose-Hubbard interaction is much larger
than all other energy scales in the problem.  This limit has the advantage of
lending itself to exact calculations, through mapping to a fermionic system
via the Jordan-Wigner (JW) transformation.  Introducing the fermionic
operators $c_{n}$ (with $\left\{ c_{n},c_{n'}^{\dagger}\right\}
=\delta_{n,n'}$) the mapping
\begin{equation}
b_{n}=S(0,n-1)c_{n},\label{eq:jw-definition}
\end{equation}
where $S(n,n')=\prod_{j=n}^{n'}\left(1-2n_{j}\right)$ is the JW string
operator, reproduces the HCB commutation relations. After the JW mapping
the fermionic Hamiltonian can be obtained from Eq.~\eqref{eq:hamiltonian-definition}
replacing $b_{n}\to c_{n}$. The bosonic one-body density matrix 
\begin{eqnarray}
\rho_{nn'}^{B} & = & \av{b_{n}^{\dagger}b_{n'}}
\end{eqnarray}
can be computed from the fermionic one, $\rho_{nn'}^{F}=\av{c_{n}^{\dagger}c_{n'}}$, using the
approach described, e.g., in Ref.~\cite{Rigol_2005}.

For uniform 1D bosons at zero temperature, the occupancy of the lowest
single-particle state (with momentum $k=0$) scales as
$n_{k=0}\propto\sqrt{N_{b}}$.  While Bose Einstein condensation (BEC) is not
strictly present in 1D, the off-diagonal elements of the density matrix still
develop a algebraic decay $\rho_{nn'}^{B}\propto\frac{1}{\abs{n-n'}^{1/2}}$ as
$\abs{n-n'}\to\infty$ in the thermodynamic limit.  In spatially inhomogeneous
situations, the quantity analogous to $n_{k=0}$ is the largest eigenvalue
$\lambda_0$ of the single-particle density matrix.  The eigenvalues are
referred to as occupation numbers of \emph{natural orbitals}
\cite{Penrose_1956,Leggett_2001,Rigol_2005}.  The natural orbitals are the
corresponding eigenvectors:
\[
\sum_{j}\rho_{ij}^{B}\Phi_{j}^{n}=\lambda_{n}\Phi_{j}^{n},
\]
with $\lambda_{0}\geq\lambda_{1}\geq...$.  Quasi-condensation is signaled by
the behavior $\lambda_{0}\propto\sqrt{N_{b}}$.  Since we are dealing with
explicitly non-uniform systems, we choose to use this language (rather than
momentum occupancies) in order to describe the presence or absence of
quasi-condensation.

We consider different types of quasi-periodic potentials.  The
simplest is a single-frequency cosine periodic potential with an
irrational wavevector relative to the lattice spacing, generally known
as the Aubry-Andre potential \cite{Aubry_1980}:
\begin{equation}
V(n)=V_{1}\cos\left(2\pi q_{1}x_{n}\right);  
\quad\mathrm{with}\quad q_{1}=\frac{\sqrt{5}-1}{2}a^{-1}
\, .
\tag*{[AA]}
\end{equation}
Here $x_{n}=an$ with $a$ the lattice constant.  The single-particle
eigenstates of the AA model are all localized for $V_{1}>2$ and all
extended for $V_{1}<2$.  In this paper we are interested in extended
Aubry-Andre models where the single-particle spectra have more
intricate structure, in particular mobility edges.  We introduce
two potentials with two frequencies:
\begin{equation}
V(n) = V_{1}\cos\left(2{\pi}q_{1}x_{n}\right) 
+V_{2}\cos\left(2{\pi}q_{2}x_{n}\right) . 
\tag*{[EAA-1,2]}
\end{equation}
The first model has the parameters
\begin{equation}
V_{2}= \tfrac{1}{6}V_{1}, \; q_{1}=\frac{\sqrt{5}-1}{2}a^{-1},\, q_{2}=3q_{1}
\tag*{[EAA-1]}
\end{equation}
and the other has parameters 
\begin{equation}
V_{2}= \tfrac{1}{3}V_{1}, \; q_{1}=\frac{0.7}{2\pi}a^{-1},\, q_{2}=2q_{1}
\, .
\tag*{[EAA-2]}
\end{equation}
We also consider another type of modification of the AA potential:
\begin{equation}
\begin{split}
V(n)=V_{1}\cos\left(2\pi q_{1}x_{n}^{\alpha_{d}}\right);  
\\  q_{1}=\frac{\sqrt{5}-1}{2}a^{-1},
\; \alpha_{d}=0.7 \, .
\end{split}
\tag*{[EAA']}
\end{equation}
These models EAA-1,2 \cite{Soukoulis_1982, Hiramoto} and
EAA' \cite{Sarma_1988} are known examples from a large class of extended
Aubry-Andre models whose single-particle spectrum possess mobility edges.
%

As a function of the ratio $V_{1}/J$ these models present the same
qualitative features: for small values of $V_{1}/J$ all the single
particle eigenvectors are extended in real space and for large values
of this ratio all the eigenstates are localized. For an intermediate
range of $V_{1}/J$ both localized and delocalized eigenstates
coexist. The single particle spectrum is organized in rather well
defined regions corresponding to localized or extended eigenvectors
separated by sharply defined mobility-edges. The generic behavior of
the localized-extended transition is illustrated in Fig.\ \ref{fig:1-1}
where the inverse participation ratio
$\text{IPR}\left(\psi\right)=\frac{\sum_{l}\abs{\psi_{l}}^{4}}{\left[\sum_{l}\abs{\psi_{l}}^{2}\right]^{2}}$
is plotted for the single-particle eigenstates of of $H$.
Note that for the AA model there are no mobility edges in the
spectrum, instead, the localized-extended transition occurs for all
eigenstates simultaneously for the same value of $V_{1}/J$.  This
non-generic feature is due to the self-duality of the model
\cite{Aubry_1980}.  Fig.\ \ref{fig:1-2} shows the single-particle spectrum as a
function of $V_{1}/J$ and illustrates the difference between
(exponentially) localized and extended states.

Fig.\ \ref{fig:1-1} illustrates that the models EAA-1, EAA-2 and EAA'
present the same generic behavior.  We therefore present our results
mainly for one of the models (EAA-1). The expectation is that the
concepts and results emerging from this work are generically valid for
any potential for which the single-particle spectrum has mobility
edges of this generic type.

\section{Ground-state properties  \label{sec_equilibrium}}

\begin{figure}[tb]
\centering  
\includegraphics[width=0.67\columnwidth]{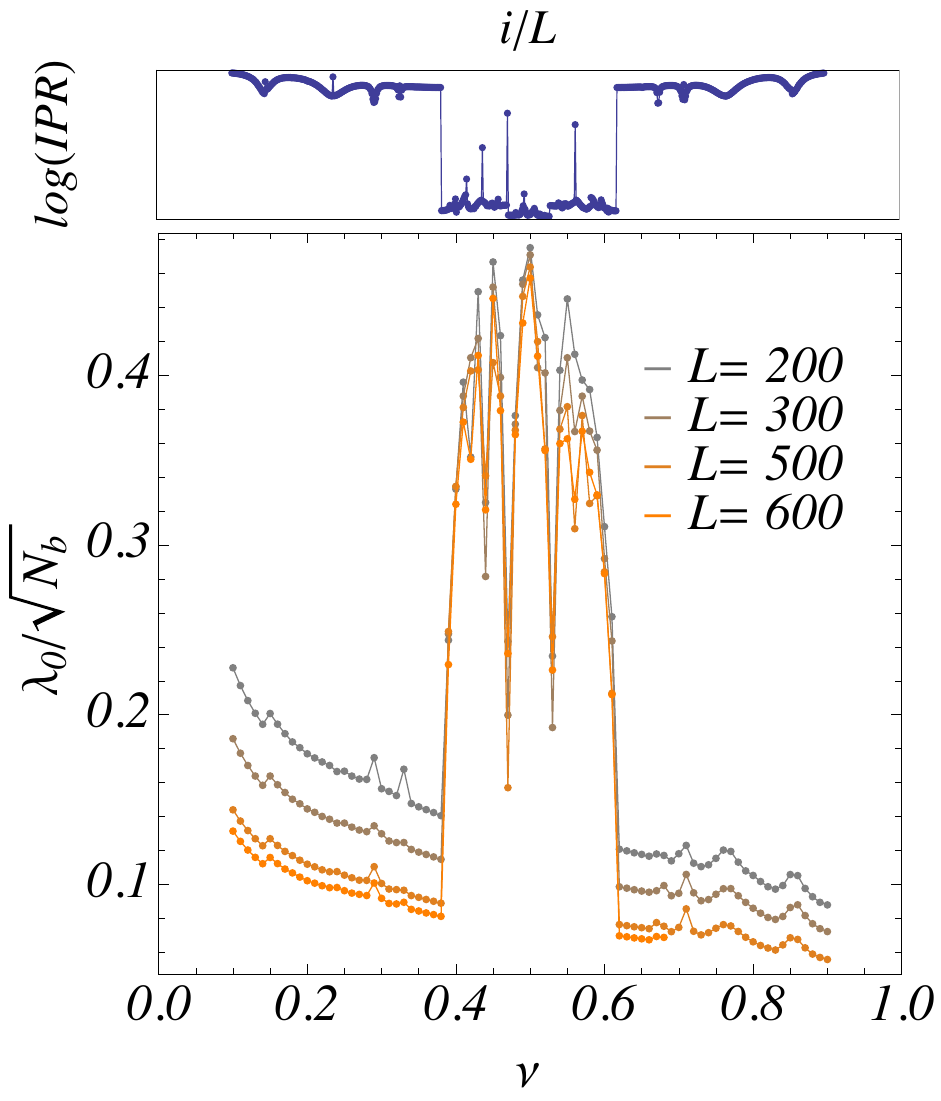}
\caption{\label{fig:2-1} Lower panel: quasi-condensate fraction
given by $\lambda_{0}/\sqrt{N_{b}}$ as a function of the filling fraction $\nu=N_{b}/L$.
($V_1=2.5J$, EAA-1.)  Quasi-condensation is observed when the chemical potential
$\mu\left(\nu\right)$ lies in a spectral region with extended states (in this case for $\nu$ around
half-filling, as seen from the top panel).  Otherwise $\lambda_{0}/\sqrt{N_{b}}$ vanishes in the
thermodynamic limit $N_{b}\to\infty$.  Upper panel: inverse participation ratio against eigenstate
index.
}
\end{figure}

In this Section we will present results characterizing the ground state properties of our system, as
a function of the filling fraction $\nu=N_{b}/L$.  The energy of the last filled JW fermionic level
is the chemical potential of the system, $\mu\left(\nu\right)$.  Our major result is that the system
behaves like an insulator or a quasi-condensate, depending on whether $\mu\left(\nu\right)$ lies in
an energy region of localized or extended single-particle states, which we will denote respectively
by $\Sigma_{l}$ and $\Sigma_{e}$.

In order to characterize the ground state we consider the behavior of the natural orbitals
occupation, the characteristic decay of the off-diagonal density matrix components, and the
entanglement entropy of a subsystem.  For definiteness we display results for the EEA-1 model with
$V_{1}=5/2J$, where the single-particle spectrum has a well-defined intermediate delocalized region
($\Sigma_{e}$) separated by mobility edges from higher and lower energy regions of localized
states ($\Sigma_{l}$).

When all the single-particle eigenstates are localized (1D random potential or AA model with
$V_{1}>2$), $\lambda_{0}/\sqrt{N_{b}}\to0$ \cite{Nessi_2011, Cai2}, in contrast to the
quasi-condensate behavior $\lambda_{0}\sim\sqrt{N_{b}}$.  In addition, in the localized (insulating)
case, the off-diagonal correlations decay exponentially $\rho_{nn'}^{B}\propto e^{-\abs{n-n'}/\xi}$ 
\cite{Nessi_2011, Cai2}, in contrast to the quasi-condensate behavior  $\rho_{nn'}^{B}\propto\frac{1}{\abs{n-n'}^{1/2}}$.   
In the AA model the two behaviors are seen for $V_{1}>2$ an $V_{1}<2$ at any filling
fraction \cite{Cai2}.  We will show that when the single-particle spectrum has mobility edges,
either behavior can appear, depending on the filling.

Fig.~\ref{fig:2-1} shows the rescaled lowest natural orbital occupation $\lambda_{0}/\sqrt{N_{b}}$
as a function of the filling fraction.  The finite size scaling shows that if
$\mu\left(\nu\right)\in\Sigma_{l}$, $\lambda_{0}/\sqrt{N_{b}}\to0$ in the thermodynamic limit, while
$\lambda_{0}/\sqrt{N_{b}}$ approaches a finite value for $\mu\left(\nu\right)\in\Sigma_{e}$.

\begin{figure}[tb]
\centering
\includegraphics[width=0.8\columnwidth]{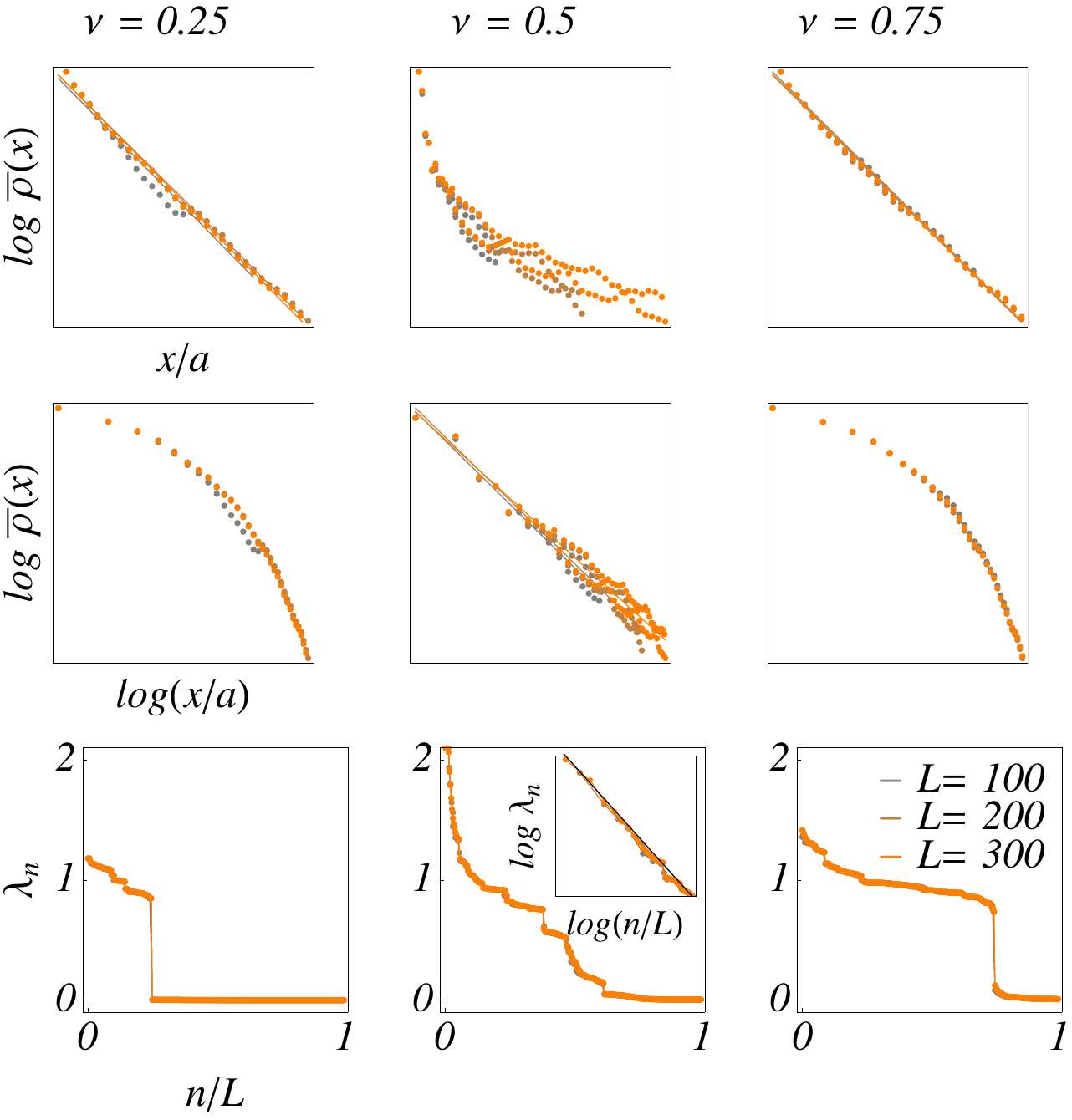}
\caption{ \label{fig:2-2}
Decay of the off-diagonal elements of $\rho^{B}$ for three different filling fractions in log-linear
(upper row) and log-log (middle row) scales.  ($V_1=2.5J$, EAA-1.)  Data points are shown for three
different system sizes $L$ to indicate the degree of convergence.  The straight lines are fits.  For
$\nu=0.25$ and $0.75$ [$\mu\left(\nu\right)\in\Sigma_{l}$, no quasi-condensation], the decay is seen
to be exponential; $\bar{\rho}^{B}\left(x\right)\propto A\, e^{-\abs x/\xi}$. For $\nu=0.5$ the decay
is well approximated by power law $\bar{\rho}^{B}\left(x\right)\propto a\,\abs x^{-\alpha}$ with
$\alpha\simeq0.59$.  Lowest row: the natural orbital occupations $\lambda_{n}$. For $\nu=0.25$ and
$0.75$ the distribution is step-like. For $\nu=0.5$ it diverges for $n{\to}0$ in the thermodynamic
limit.  This can be seen in the inset where the black line corresponds to a divergence of the form
$\lambda_{n}\propto c\left(n/L\right)^{-\kappa}$ with $\kappa\simeq0.3$.
}
\end{figure}

\begin{figure}[tb]
\centering{}%
\begin{tabular}{ll}
(a) & (b)\tabularnewline
\begin{tabular}{c}
\includegraphics[width=0.4\columnwidth]{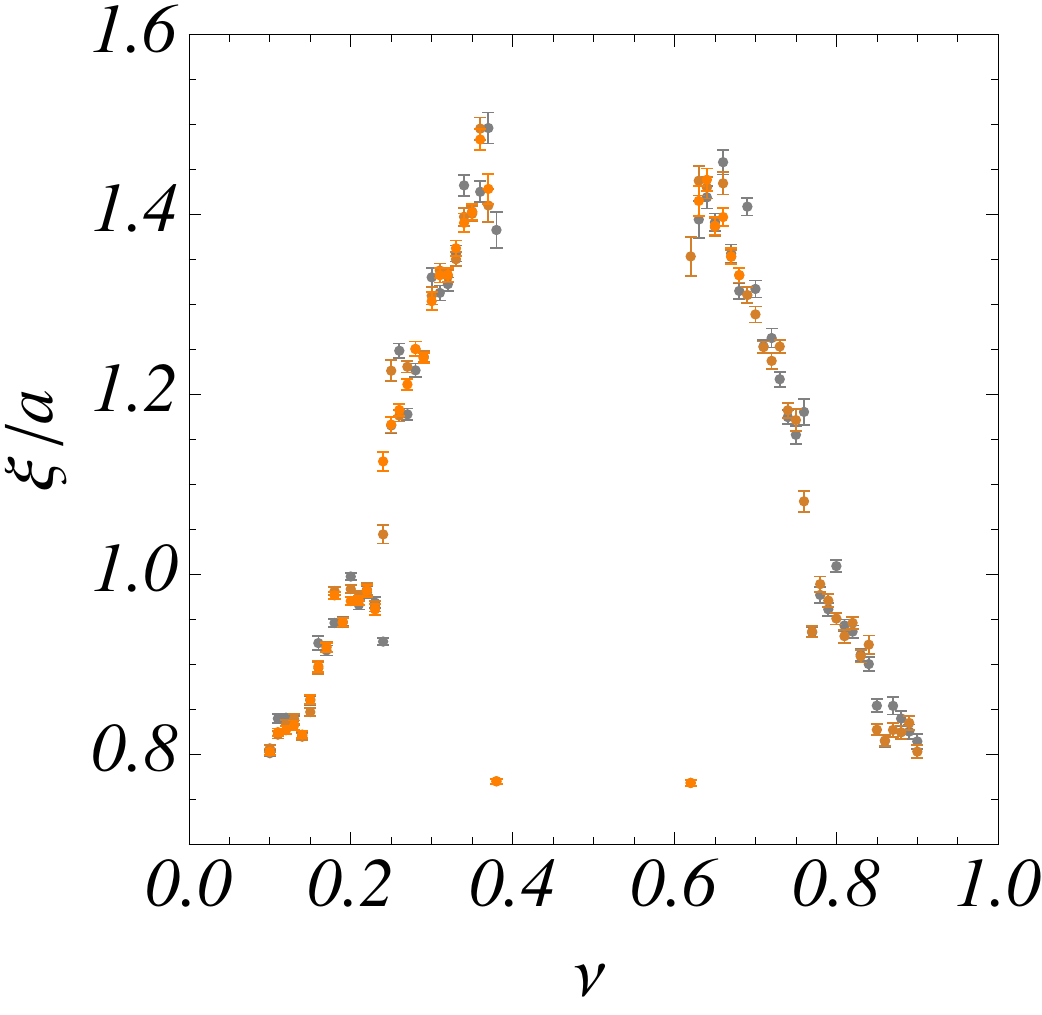}\tabularnewline
\end{tabular} & %
\begin{tabular}{c}
\includegraphics[width=0.4\columnwidth]{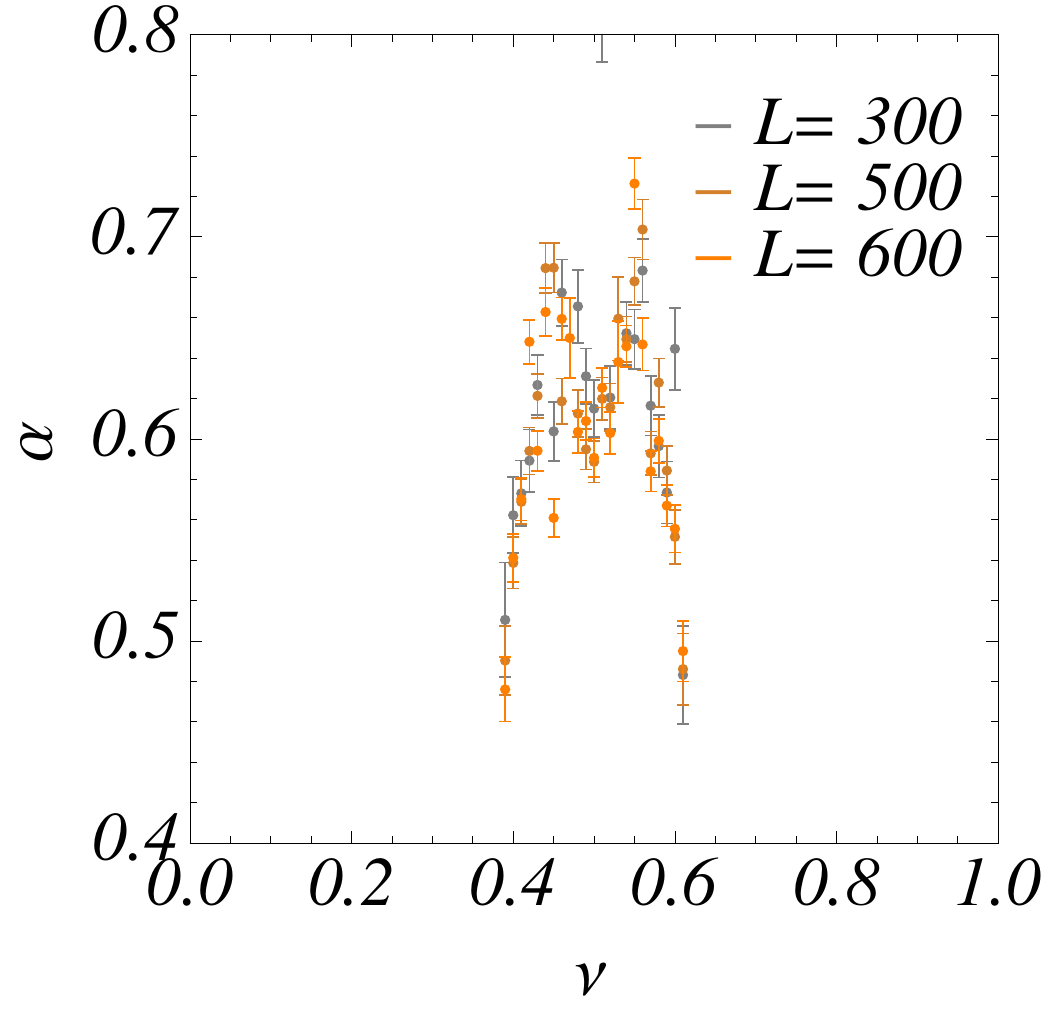}\tabularnewline
\end{tabular}\tabularnewline
\end{tabular}
\caption{  \label{fig:2-3}
($V_1=2.5J$, EAA-1.)  (a) Lengthscale $\xi$ of exponential decay of the off-diagonal elements of
$\rho^{B}$, obtained by fitting with an exponential form $\bar{\rho}^{B}\left(x\right)\propto
e^{-\abs x/\xi}$ (see text).  The error bars are given by the fit to the numerical data.  For the
$\nu$ values where quasi-condensation occurs, the behavior is algebraic (Fig.~\ref{fig:2-2}) and no
$\xi$ values are shown.
(b) Power-law exponent for off-diagonal decay $\bar{\rho}^{B}\left(x\right)\propto \abs x^{-\alpha}$
in the region where quasi-condensation is present.  
%
%
}
\end{figure}

The stark difference between the $\mu\left(\nu\right)\in\Sigma_{l}$ and
$\mu\left(\nu\right)\in\Sigma_{e}$ cases is also observed in the off-diagonal elements of the
averaged one-body density matrix $\bar{\rho}^{B}\left(x_{j}\right)
= \frac{1}{L}\sum_{i}^{L}\abs{\rho_{i,i+j}^{B}}$ (see Figs.\ \ref{fig:2-2}, \ref{fig:2-3}).  Here the
average is taken to smoothen oscillations introduced by the quasi-periodic potential.  For
$\mu\left(\nu\right)\in\Sigma_{l}$ the decay is exponential $\bar{\rho}^{B}\left(x\right)\propto
e^{-\abs x/\xi}$ whether for $\mu\left(\nu\right)\in\Sigma_{e}$ one observes an algebraic behavior
$\bar{\rho}^{B}\left(x\right)\propto \abs x^{-\alpha}$.  We found the exponent $\alpha$ to be in the
range $\alpha\in\left[0.5,0.7\right]$ for our system sizes of $\mathcal{O}(10^3)$, for EAA-1, EAA-2,
EAA' models.  This is higher than in the bare tight-binding lattice ($V_{1}=0$) case, for which
$\alpha=1/2$ \cite{Rigol_2005}.  It is currently not known whether $\alpha$ would converge to $1/2$
in the thermodynamic limit in multi-chromatic potentials with fillings corresponding to
$\mu\left(\nu\right)\in\Sigma_{e}$.

The occupation of the natural orbitals, displayed in Fig.\ref{fig:2-2} (lower panel), shows that for
$\mu\left(\nu\right)\in\Sigma_{l}$ the distribution of the $\lambda_{n}$'s is step-like, contrasting
with the smooth decay for the $\mu\left(\nu\right)\in\Sigma_{e}$ case. 
Step-like features are usually assigned to fermionic distributions.  Here the natural orbitals are
localized \cite{Nessi_2011} and the bosons behave much like in a fermionic system as no localized
orbital can accommodate more than one HCB.

Another striking difference between $\mu\left(\nu\right)\in\Sigma_{l}$ and
$\mu\left(\nu\right)\in\Sigma_{e}$ cases is observed in the behavior of $\lambda_{n}$ at $n{\to}0$.
For $\mu\left(\nu\right)\in\Sigma_{e}$ one observes a divergence
$\lambda_{n}\propto \left(n/L\right)^{-\kappa}$.  For the example displayed in Fig.\ref{fig:2-2},
$\kappa\simeq0.3$.  However this exponent does depends on $\nu$ and on the specific model of
disorder and can even be larger the value $\kappa=1/2$ obtained for $V_{1,2}=0$.  For
$\mu\left(\nu\right)\in\Sigma_{l}$ no divergence is observed.

\begin{figure}[tb]
\centering
\includegraphics[width=0.9\columnwidth]{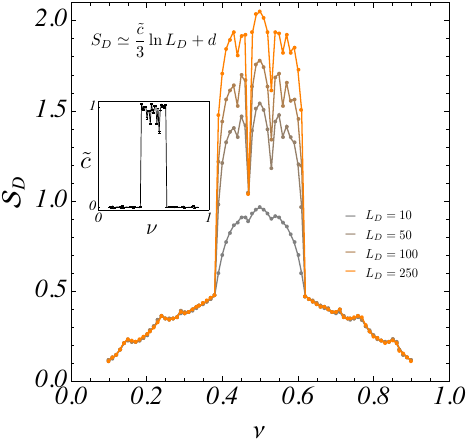}
\caption{ \label{fig:2-4} 
Ground state entanglement entropy $S_D$ of a subsystem $D$, averaged over the
central position of the subsystem, for various subsystem sizes $L_{D}$.
($L=1000$.)  
For the $\nu$ values where quasi-condensation does not occur [$\mu\left(\nu\right)\in\Sigma_{l}$],
$S_{D}$ saturates to a constant value with increasing $L_{D}$.  In the
$\mu\left(\nu\right)\in\Sigma_{e}$ region, $S_{D}$ grows logarithmically with $L_{D}$.  A
logarithmic fit (inset) shows that in this region the coefficient $\tilde{c}$ approaches the value
$c=1$ which is the exact result for the case where no quasi-periodic potential is present.
}
\end{figure}

Finally, we consider scaling of the GS entanglement entropy of a subsystem $D$ as a function of the
subsystem size $L_{D}$.  The entanglement entropy is defined as
$S_{D}=-\tr_{D}\left[\hat{\rho}_{D}\ln\hat{\rho}_{D}\right]$, where
$\hat{\rho}_{D}=\tr_{\bar{D}}\left[\hat{\rho}\right]$ and $\hat{\rho}=\ket{\Psi_{0}}\bra{\Psi_{0}}$
are many-body density matrices and $\tr_{D}$ ($\tr_{\bar{D}}$) denote the trace over the degrees of
freedom in $D$ (in the complement of $D$).  For HCB the expression for the entanglement entropy
simplifies \cite{Vidal_2003} to $S_{D}=-\sum_{i}\nu_{i}\ln\nu_{i}+(1-\nu_{i})\ln(1-\nu_{i})$ where
the $\nu_{i}$'s are the eigenvalues of the two-body density matrix of the JW fermions restricted to
the subsystem $D$, $\rho_{ij}^{F}$ with $i,j\in D$. For $\mu\left(\nu\right)\in\Sigma_{l}$, $S_{D}$
saturates as $L_{D}\to\infty$. However for $\mu\left(\nu\right)\in\Sigma_{e}$ it behaves as
$ $$S_{D}\simeq \frac{\tilde{c}}{3}\ln L_{D}$ for large subsystem sizes $L_{D}$ with a prefactor
$\tilde{c}\simeq1$. This represents the well-known logarithmic correction to the ``area law'' in
gapless one-dimensional systems \cite{Vidal_2003, Calabrese_2004}, with the prefactor given by the
central charge $c=1$ in this case.  
Once again, the bosonic system behaves like a gapless quasicondensate or like a gapped insulator
depending on the location of the ``Fermi energy'' of the corresponding free-fermion system.

\section{Expansion dynamics \label{sec_dynamics}}

\begin{figure*}[tb]
\centering{}%
\begin{tabular}{c}
\includegraphics[width=1.7\columnwidth]{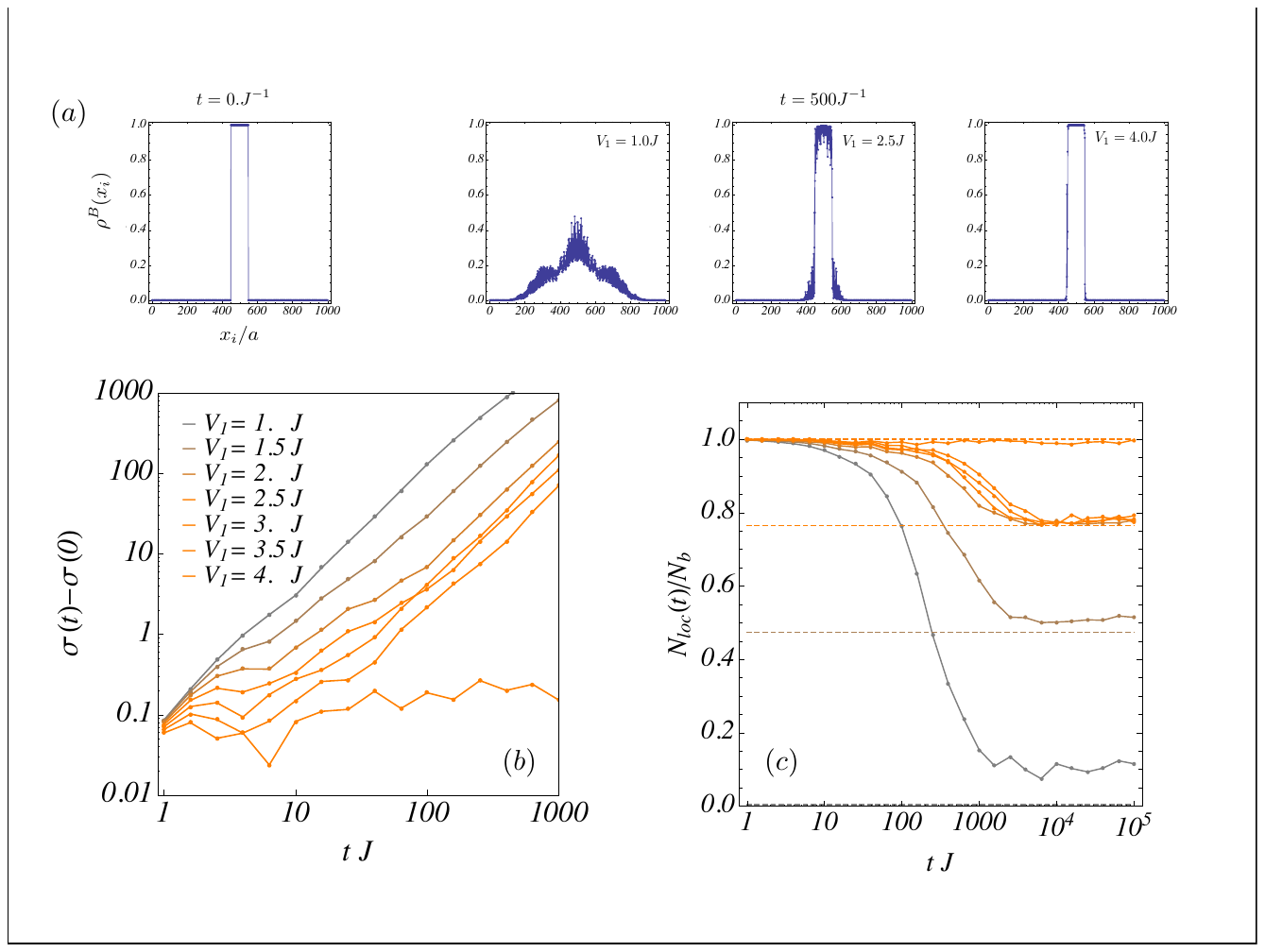}\tabularnewline
\end{tabular}\caption{  \label{fig:3}
Expansion dynamics of bosons initially clustered in the center of the lattice
as shown in top left.  ($N_{b}=100,\, L=1000$.)
In the other (a) panels, the density profile at $t=500J^{-1}$ is shown for
$V_{1}=1J$, $2.5J$, $4J$ (extended, mobility-edge, localized regimes).  
(b) Root-mean-square width of the expanding cloud.  When there is a fraction
of extended states and since they have a non-zero overlap with the initial
state, the evolution is always ballistic for large times
$\sigma\left(t\right)\propto v\, t$, nevertheless the pre-factor $v$ gets
smaller with increasing $V_{1}$ as the fraction of localized eigenstates
increases. When all the single particle states are localized the cloud width
saturates for large times. (c) Time evolution of the number of bosons in the
central region $S$ (see text).  The dashed lines give the fraction of
delocalized eigenstates in the single-particle spectrum. }
\end{figure*}

In this section we address the non-equilibrium expansion dynamics of an initially trapped cloud of
bosons after the trapping potential is turned off at time $t=0$.
The motivation for studying this type of dynamical protocol is that expansion dynamics is
particularly sensitive to the localization properties of the underlying single-particle states:
particles will mostly fly off or mostly remain in the initial region depending on whether the
initial overlap is dominantly with extended or localized single-particle states.
This type of physics has also been explored in two recent experiments, which use higher-dimensional
disordered potentials having mobility edges \cite{Kondov_2011, Jendrzejewski_2012}.

The initial state ($t<0$) is the ground state of the Hamiltonian \eqref{eq:hamiltonian-definition}
with an added harmonic trapping potential, i.e.,
\begin{equation}
H(t) ~=~ H  ~+~  \Theta(-t)\sum_{n} W(n) b_{n}^{\dagger}b_{n}  
\label{eq:hamiltonian_plus_trap}
\end{equation}
where $\Theta(...)$ is the Heaviside step function and
\begin{equation}
W(n)=W(x_{n}-x_{n_0})^{2} ,
\end{equation}
$n_0$ being a site near the center of the chain.  We monitor the cloud size
$\sigma\left(t\right)= \left[\av{\hat{x}_{i}^{2}(t)}-\av{\hat{x}_{i}(t)}^{2}\right]^{1/2}$ and the
fraction of localized atoms $N_{\text{loc}}(t)/N_{b}$ as a function of time. Here,
$N_{\text{loc}}(t)$ is defined as the number of atoms remaining in the support $S$ of the initial
density distribution, $S=\left\{ i:\av{\hat{n}_{i}(t=0)}\neq0\right\} $.

In the presence of a trapping potential and for $V_{1,2}=0$, a single dimensionless parameter
$\tilde{\rho}=N_{b}\sqrt{W/J}$ controls the behavior of the atomic density in the large-$N_{b}$
limit \cite{Rigol_2005}. This means that $\rho_{nn}^{B}\to
g\left(n\tilde{\rho}/N_{b};\tilde{\rho}\right)/N_{b}$ as $N_{b}\to\infty$, where
$g\left(y;\tilde{\rho}\right)$ is the normalized density distribution.  For
$\tilde{\rho}\simeq2.6-2.7$ a Mott insulator region builds up in the middle of the trap, where the
density is pinned to unity.

As in the last section we focus on the model EAA-1.  We study the dynamics for different values of
$V_{1}$ and $\tilde{\rho}$.  We separate the case $\tilde{\rho}\to\infty$, for which a simple
interpretation of the results can be given, from the finite $\tilde{\rho}$ that leads to more
complex dynamics.

\subsection{Large $\tilde{\rho}$}

We start by addressing the large $\tilde{\rho}$ limit where, in the initial density distribution,
all the atoms are in the Mott phase.  In other words, in the initial state a central block of the
chain is completely filled while the rest of the chain is empty.  Fig.\ref{fig:3}(a) shows snapshots
of the time evolution of the initial density profile (left panel) for different values of $V_{1}$.

Figs.\ \ref{fig:3}(b,c) display the time evolution of the cloud size and the fraction of localized
particles. For $V_{1}\lesssim3.8J$, $\sigma\left(t\right)\propto v\, t$ (for $t\to\infty$), which
means that at least some part of the cloud spreads ballistically. The spreading velocity $v$ is a
decreasing function of $V_{1}$. The threshold value $V_{1}\simeq3.8t$ is that above which there are
no more extended eigenstates in the spectrum, as can be seen from Fig.\ref{fig:1-1}(b).  For
$V_{1}\gtrsim3.8J$, $\sigma\left(t\right)$ saturates and the width of the cloud remains bounded for
large times.

From Fig.\ \ref{fig:3}(c) one can see that, at long times, the fraction of particles that remain
within the region $S$ is given by the fraction of extended states in the spectrum (horizontal dotted
lines). This may be understood by considering the evolution in the eigenbasis of the non-trapped
Hamiltonian ($W_{n}=0$) with which the system evolves for $t>0$.  
In the asymptotic long time limit the fraction of delocalized atoms is proportional to the overlap
of the initial state with the extended eigenstates of the non-trapped Hamiltonian.
If the number of initially occupied sites is not too small, the initial state has approximately an
equal overlap with all localized and extended eigenstates.  The fraction of particles localized in
$S$ should thus equal the fraction of extended eigenstates in the spectrum.

\subsection{Arbitrary $\tilde{\rho}$}

\begin{figure}[tb]
\centering{}%
\begin{tabular}{c}
\begin{tabular}{l}
\tabularnewline
\includegraphics[width=0.9\columnwidth]{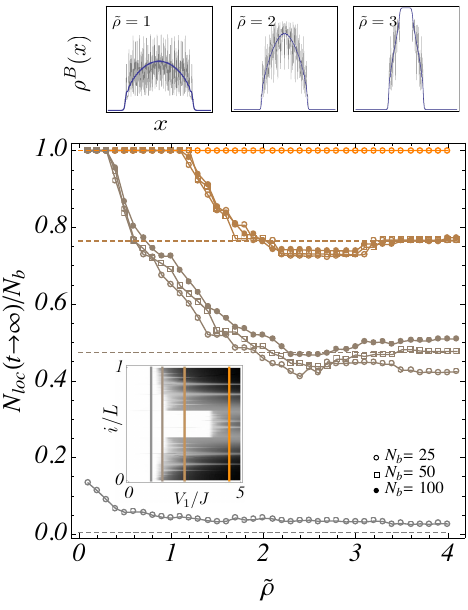}\tabularnewline
\end{tabular}\tabularnewline
\end{tabular}\caption{\label{fig:toftrap} Main panel: fraction of bosons remaining localized
for asymptotically long times after switching off the trapping potential, as a function of parameter
$\tilde{\rho}$.  ($L=2000$, EAA-1, various $V_{1}$ and $N_{b}$.) For large $\tilde{\rho}$ the
asymptotic value is given by the fraction of delocalized eigenstates in the spectrum (horizontal
dashed lines). For small $\tilde{\rho}$ all the bosons remain localized after releasing the
trap. The inset displays the IPR (as in Fig.\ \ref{fig:1-1}); the colored lines correspond to the
values of $V_{1}/J$ of the main panel.  Upper panels: the initial density profile is displayed for
$V_{1}/J=0$ (blue) and for $V_{1}/J=1$ (gray), for different values of $\tilde{\rho}$.
}
\end{figure}

Next, we address the time evolution starting from the ground state of a harmonic trapping potential
characterized by a finite value of $\tilde{\rho}$. Fig.\ \ref{fig:toftrap} shows the asymptotic
long-time value of the fraction of atoms localized in $S$ as a function of $\tilde{\rho}$ for
different values of $V_{1}$ and for $N_{b}=25,50,100$.
The convergence of our results for increasing $N_{b}$ shows that the physics of the large-$N_{b}$
limit is well-represented in our numerical simulations and also that, for these $N_{b}$ values,
effects of the location of the quasiperiodic potential is sufficiently averaged over.

For $\tilde{\rho}=0$ (no trap) and $V_{1}\gtrsim1.2J$ all the atoms are localized as the ground
state is obtained by filling the first $N_{b}$ single-particle levels, which are all localized for
the EAA-1 potential in this filling region. $ $For small $\tilde{\rho}$ (very shallow trap) one can
still find $N_{b}$ localized single-particle eigenstates within the trapping length
$\ell=N_{b}a/\tilde{\rho}$.  This explains why all bosons remain localized for small but nonzero
values of $\tilde{\rho}$.

However, as $\tilde{\rho}$ increases, the number of localized states available within a region of
size $\ell$ starts to be smaller than $N_{b}$ and the overlap with delocalized states is then
finite. For large $\tilde{\rho}$ the results of the last section are recovered and the fraction of
localized atoms equals the fraction of localized states in the spectrum.

\section{Summary \& Discussion  \label{sec_discussion}}

We have presented a study of a strongly interacting many-boson system in extended Aubry-Andre models
for which the single-particle spectrum displays mobility edges, i.e., in situations where both
extended and localized states are present in regions ($\Sigma_{e}$ and $\Sigma_{l}$) of the
single-particle spectrum.  The bosonic system was treated by mapping to free fermions, so that
numerically exact calculations are possible for relatively large systems.  We have shown through
non-equilibrium calculations that expansion dynamics can be used as a probe of the sizes of
$\Sigma_{e}$ and $\Sigma_{l}$ regions in the single-particle spectrum.

For ground-state properties, the most striking result is that the properties of the many-body system
(insulating or quasi-condensate) is determined solely by the location of the chemical potential
$\mu(\nu)$, which has the interpretation of being the Fermi energy of the corresponding free-fermion
system.  We have illustrated this by taking a representative model where, in a range of potential
strengths, increasing the filling fraction $\nu$ takes the chemical potential from a
$\mu\left(\nu\right)\in\Sigma_{l}$ region at low densities, through a region of
$\mu\left(\nu\right)\in\Sigma_{e}$ region at intermediate $\nu$, to a
$\mu\left(\nu\right)\in\Sigma_{l}$ region at high densities.  For different fillings $\nu$, we
presented the occupations of natural orbitals ($\lambda_{n}$, particularly the quasi-condensate
occupancy $\lambda_0$), the off-diagonal decay of the single-particle bosonic density matrix, and
the entanglement entropy scaling.  These observables all show characteristics of gapped insulators in
fillings for which $\mu\left(\nu\right)\in\Sigma_{l}$, and show characteristics of gapless
quasi-condensates at $\nu$ values for which $\mu\left(\nu\right)\in\Sigma_{e}$.

At present we lack a simple explanation for this result that the location (relative to regions of
the single-particle spectrum) of the Fermi surface of the corresponding free-fermion system
determines off-diagonal properties of strongly interacting bosons.  While properties of fermionic
systems are often described by features near the Fermi surface, the properties we have shown are
bosonic rather than fermionic.  Also, it is rather unintuitive that, when
$\mu\left(\nu\right)\in\Sigma_{l}$, no quasi-condensate properties are seen even if a sizable
fraction of Jordan-Wigner fermions occupy extended states.

Ref.~\cite{Prodan_2010} has observed, for another case (free fermions in disordered 2D Chern
insulators) where the single-particle states has different natures in different parts of the
spectrum, that certain properties depend solely on the nature of the eigenstates at the location of
the Fermi surface.  This loosely similar observation in a very different context suggests that the
situation is generic for systems where the single-particle spectrum contains regions of different
nature.

These results open up several questions that deserve to be addressed in future research.  Most
prominently, one would like to know if this dependence on the location of the filling fraction
extends far beyond the hard-core limit for bosons with strong but finite interactions (e.g.,
Bose-Hubbard chain with finite $U$).  This would be particularly intriguing because such a system
does not map exactly to free fermions, so that a simple picture of filling up the fermi sea does not
hold.
More speculatively, one might also wonder if bosons in higher dimensional lattices with mobility
edges also have properties determined solely by the location of the chemical potential, since in
higher dimensions there is no connection to a fermionic picture.
Another interesting direction would be to study finite-temperature properties.  Finite temperatures
might allow the system to explore the $\Sigma_{l(e)}$ region even when the chemical potential is in
the $\Sigma_{e(l)}$ region, if the chemical potential is sufficiently close to a mobility edge
separating the two regions.

\end{document}